\documentclass{ws-mpla1}
\usepackage{graphicx}
\usepackage{color} 
\usepackage{hyperref}
\usepackage{url}
\usepackage{amssymb,amsmath}
\usepackage{amstext}
\usepackage{amscd}
\usepackage{cite}
\usepackage{yhmath,mathrsfs,graphicx}
\usepackage{amstext}

\def\sideremark#1{\ifvmode\leavevmode\fi\vadjust{\vbox to0pt{\vss
 \hbox to 0pt{\hskip\hsize\hskip1em
 \vbox{\hsize3cm\tiny\raggedright\pretolerance10000
  \noindent #1\hfill}\hss}\vbox to8pt{\vfil}\vss}}}

\DeclareMathAlphabet{\mathpzc}{OT1}{pzc}{m}{it}

\begin{document}

\markboth{S. Deser, K. Izumi, Y. C. Ong, A. Waldron}
{Problems of Massive and Bimetric Gravities}

\catchline{}{}{}{}{}

\title{Problems of Massive Gravities}

\author{S. Deser}

\address{
Walter Burke Institute for Theoretical Physics, Caltech, Pasadena, CA 91125; \\
     Physics Department, Brandeis University, Waltham, MA 02454, USA. \\
{\tt deser@brandeis.edu}}

\author{K. Izumi}

\address{Leung Center for Cosmology and Particle Astrophysics, \\
National Taiwan University, Taipei 10617, Taiwan. \\
{\tt  izumi@phys.ntu.edu.tw}
}

\author{Y. C. Ong}

\address{
Nordita, KTH Royal Institute of Technology and Stockholm University,\\ Roslagstullsbacken 23,
SE-106 91 Stockholm, Sweden.\\
{\tt yenchin.ong@nordita.org}
}

\author{A. Waldron}

\address{Department of Mathematics, University of California, Davis, CA 95616, USA.\\
 {\tt wally@math.ucdavis.edu}
}

\maketitle


\begin{abstract}
The method of characteristics is a key tool for studying consistency of equations of motion; it allows 
 issues such as predictability, maximal propagation speed, superluminality, unitarity and acausality to be addressed without requiring explicit solutions.  
We review this method and its  application to massive gravity theories to 
 show the limitations of these models' physical viability: Among their problems are loss
of unique evolution, superluminal signals, matter coupling inconsistencies and micro-acausality (propagation of signals around local closed timelike/causal curves).
We extend previous no-go results to the entire  three-parameter range of massive gravity theories.
 It is also  argued that bimetric models   suffer a similar fate.
\end{abstract}


\section{Introduction}	

Among the many recent attempts to generalize/modify General Relativity (GR), those that break its general covariance and make its range finite, by adding a mass term to the Einstein action, are perhaps the most active current contenders. These massive gravity (mGR) models are themselves of two types: fiducial massive gravity (fmGR) proceeds 
by embedding the dynamical ``Einstein'' system in a fixed spacetime background. Instead,  the so-called bimetric ($f$-$g$ or biGR) models also make the second metric dynamical; the system now has a single, shared, diffeomorphism invariance. While both directions were introduced long ago~\cite{Salam,Zumino}, their recent revival is due to the (partial) resolution of an earlier, fatal, flaw~\cite{BD}: mGR models generically propagate a sixth, ghostlike, mode in contrast with  the five physical degrees of freedom (DoF) of their linearized, Fierz--Pauli (FP)~\cite{FP} counterparts. 
This terminated interest in bimetric and fmGR models for four decades. The resolution of this no-go impasse  was based first on effective field theory reasoning that used  a decoupling limit
(large Planck mass, small graviton mass-squared, and a constant product of the two) to study  distinguished mass terms depending on~$(g_{\mu\nu},\bar g_{\mu\nu})$, identifying
just  three ``ghost-free'', combinations~\cite{dRGT1} (see also the reviews~\cite{HinterbichlerReview,deRham:2014zqa}). Later, this conjecture was confirmed (at least---overoptimistically as we shall see---assuming invertibility of certain constraints) by ADM~$3+1$ Hamiltonian techniques~\cite{HR}.
Indeed, we shall show here that in the process of removing the original ghost, new---and equally fatal---flaws arise as an unavoidable part of the ``cure'', a result that restores its unique place to Einstein's coordinate invariant long range~GR, just as the uniqueness of massless Yang-Mills theories was forced (for other---quantum---reasons). 

At their core, mGR models are described by a coupled set of quasilinear partial differential equations (PDEs). These can be studied in detail using the method of characteristics, which allows the models' predictability, maximal propagation speed, superluminality and acausality to be addressed, both for explicit solutions and in full generality. 
Since the method determines whether initial data can be propagated into the future, it also allows quantum consistency (recall the relationship between quantum field commutators and propagators~\cite{Bjorken:1965zz}) to  be  addressed; these are methods that have been employed
since the very earliest studies of massive higher spin systems~\cite{KS,VZ}. Therefore, we begin in Sec.~\ref{BRChara} with a review of the characteristic method.

Armed with this mathematical technology, we shall  tackle in detail the fmGR models. These are defined in Sec.~\ref{fmGR}. In order to study their characteristics, it is extremely propitious
to have explicit covariant expressions for their constraints. The first analysis of this type was given in~\cite{DMZ} where it was proven---covariantly---that (for at least one of the three mass terms) the model generically propagates~5~DoF.
Explicit expressions for these were given in~\cite{DW} and extended to a two-dimensional subspace of fmGR parameter space in~\cite{DSW}. The remaining, unprobed, third parameter choice remained problematic for quite some time because the dynamical Weyl tensor appeared in its would-be scalar constraint---the one responsible for
ghost-removal. The covariant proof that the last scalar constraint extended to the full parameter space was only very recently provided in~\cite{DSWZ}. That paper gives a
definitive analysis of fmGR's constraint structure; it is  briefly summarized in Sec.~\ref{fmGR}.

Characteristics for fmGR theories were first studied in~\cite{DW} where superluminalities were discovered. (Earlier work also found superluminality in the models' decoupling limit~\cite{Gr}.)
The models' full characteristic matrix was then computed, at least for a subspace of parameter space in~\cite{Izumi:2013poa,Deser:2013eua}{}; the analysis of~\cite{Deser:2013eua} then even uncovered acausalities--closed timelike curves (CTCs). The characteristic matrix for the full fmGR parameter space was finally computed in~\cite{DSWZ}. Those results are summarized in Sec.~\ref{char}, where we also use them   to give  two explicit acausality examples, one a very simple one and the second which  invalidates mGR in flat fiducial backgrounds for any combination of the three possible mass terms.
Our conclusions and further comments on the causal properties of biGR and matter coupling difficulties are given in Sec.~\ref{conc}.

Sections~\ref{fmGR} and~\ref{char} 
rely in part on work performed in collaboration with M. Sandora and G. Zahariade, originally reported in~\cite{DSW,PMbi,DSWZ}.

\section{The Method of Characteristics}\label{BRChara}

The method of characteristics is a very useful tool for analyzing field theories, both classical and quantum. For the former, it determines hypersurfaces off which the system of nonlinear PDEs determining classical evolution is not fully predictive and along which shocks propagate. In the quantum setting, the characteristic method allows quantum field commutators to be analyzed. This is because the commutator function in a quantum field theory is related to the propagator and in turn to the predictability question. In short, the characteristic method probes the kinetic structure of field theories, so we begin with a brief review (see~\cite{CH,Nester8,Ong:2013qja,DW,Izumi:2013poa,Deser:2013eua,Izumi:2013dca,Izumi:2014loa} for further details).

\subsection{Brief Review of Characteristics}

The method of characteristics allows us to  locate  hypersurfaces beyond which the evolution of a system of PDEs ceases to be unique. 
Mathematically, this is determined along the hypersurface by the vanishing of the  coefficient of the highest order derivative in the normal direction. 
It can be intuitively understood as follows:
An~$n$-th order differential equation can be generally solved once initial conditions for the~$(n-1)$-th order derivatives are specified. 
More explicitly, 
the initial conditions for~$(i+1)$-th order derivatives with~$0\le i \le n-2$ fix the evolution of~$i$-th order derivatives,
while the evolution for the~$(n-1)$-th order derivative is obtained by solving the equation for the {$n$-th}~order derivative. 
However, if its coefficient vanishes, we cannot solve the equation, and so the evolution cannot be uniquely fixed.

The detailed set-up is as follows: 
Suppose that we have a hypersurface~$\Sigma$ and a vector~$\xi^\mu$   not tangent to it
(given a metric,~$\xi^\mu$ can be chosen  normal to~$\Sigma$  for simplicity).
If the evolution equation in the direction of~$\xi^\mu$ becomes singular, the evolution cannot be unique. 
Such a hypersurface is called a characteristic hypersurface.
We can decompose a spacetime into hypersurfaces~$\Sigma$ and normal directions~$\xi^\mu$ (along the lines of  the Arnowitt-Deser-Misner (ADM) formalism\cite{ADM1},
but for now do not stipulate whether~$\xi^\mu$ is spacelike, timelike or lightlike with respect to some choice of metric). 
Now consider in this context  a  {\it quasi-linear} equation ({\it i.e.}, the highest derivative order is~$n$ and that term appears linearly) equation for a scalar field~$\phi$ 
\begin{eqnarray}\label{ap}
A^{\mu_1\cdots\mu_n}\partial_{\mu_1} \cdots \partial_{\mu_n} \phi + {\cal O}\big(\partial^{n-1}\phi\big) =0.
\end{eqnarray}
The vanishing of the coefficient of the operator ~$\xi^{\mu_1} \cdots \xi^{\mu_n}\partial_{\mu_1} \cdots \partial_{\mu_n}$ is the condition that the evolution off of~$\Sigma$ becomes singular: 
\begin{eqnarray*} 
A^{\mu_1\cdots\mu_n}\xi_{\mu_1} \cdots \xi_{\mu_n}=0\, ;
\end{eqnarray*}
this is known as  the \emph{characteristic equation}.
Observe that the form of the left hand side of this equation is the same as the first term of the left hand side of Eq.~(\ref{ap}) upon replacing~$\partial_\mu$ by~$\xi_\mu$.

\begin{figure}[tbp]
  \begin{center}
    \includegraphics[keepaspectratio=true,height=55mm]{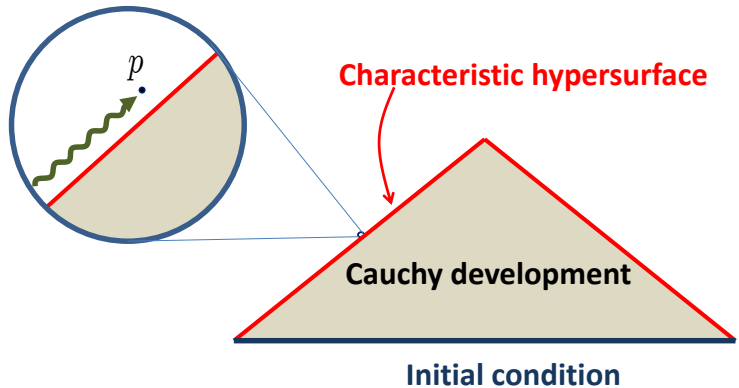}
  \end{center}
  \caption{The physics at a point~$p$ lying outside the Cauchy development of given initial datum~$S$ cannot be uniquely fixed. Propagation from outside~$S$ can affect~$p$. \label{1}}
\end{figure}

The characteristic hypersurface is of course intimately related to the \emph{maximum  propagation speed} in the theory. 
On it, the evolution of some physical variables becomes singular. This means that the hypersurface must be an edge of the 
Cauchy development from initial data~$S$. In~GR, the characteristic hypersurface is none other than the light cone emanating from~$S$ (see Fig.~\ref{1}), the speed of light being the upper bound for physical propagation. 
Consider a point~$p$ in a neighborhood of the characteristic hypersurface but outside of the Cauchy development; 
physical data at~$p$ cannot be uniquely determined given only the information on the characteristic hypersurface. 
Any point~$p$ that lies on the edge of the Cauchy development is affected by modes propagating along the characteristic hypersurface.  
Note that it is not necessary, though very convenient, to consider discontinuous modes; the discussion can be applied to any smooth (and in fact, analytic) propagation. We further comment on this point below.

\subsection{Characteristics vs. Shocks}\label{shock}

We shall now discuss the relation of characteristic hypersurfaces to shock waves (henceforth, ``shocks''). 
Consider~$n$ first order  quasi-linear equations for~$n$ (not necessarily scalar) variables~$\phi^{(a)}$:
\begin{eqnarray*}
\sum_b F_{(a,b)}^\mu \partial_\mu \phi^{(b)} +F_{(a)}=0\, ,
\end{eqnarray*}
where ``$a,b \in \left\{1,\cdots, n\right\}$" are the labels for the variables, and  
$F_{(a,b)}^\mu$ and~$F_{(a)}$ are any functions of~$\phi^{(a)}$.
As shown above, the evolution from a hypersurface becomes singular if 
\begin{eqnarray*}
\det \left[F_{(a,b)}^\mu \xi_\mu \right]=0\, ,
\label{Fab}
\end{eqnarray*}
where the vector~$\xi^\mu$ is normal to the hypersurface. 
Alternatively, we can consider a shock solution with 
 discontinuities in its  first order derivatives across a hypersurface normal to~$\xi^\mu$.
The shock  (the jump in first order derivatives) is denoted by 
\begin{eqnarray}\label{jump}
[\partial_\mu\phi^{(a)}]=\xi_\mu\tilde\phi^{(a)},
\end{eqnarray}
which implies the  junction condition  
\begin{eqnarray*}
\sum_b F_{(ab)}^\mu \xi_\mu \tilde\phi^{(b)}=0\, . \label{Fxi}
\end{eqnarray*}
Notice that the requirement for the existence of a non-trivial solution for~$\tilde\phi^{(a)}$ is that the determinant of the above matrix of coefficients is zero. 
This is exactly the characteristic matrix.
In other words, shock fronts must lie along characteristic hypersurfaces. 
Technically a shock analysis is  easier than  the more general method of characteristics, which  holds even for analytic, shock-free configurations.

\subsection{Characteristics vs. Strong Coupling}

The analysis of characteristics is completely independent of the strong coupling scale.
For instance,  consider a  scalar field action in Minkowski spacetime:
\begin{eqnarray*}
S=-\frac12 \int d^4x \left[{\alpha^2} \Big( \partial_\mu  \varphi\  \partial^\mu \varphi + m^2 \varphi^2\Big) + \, \frac\lambda 6\ \varphi^4\right].
\end{eqnarray*}
Because  the characteristic method analyzes propagation, it focusses on the  behavior of the kinetic term. 
Thus here (and in many other similar situations) at energy scales ~$E\gg m$, a good approximation is to ignore  the mass term. 
On the other hand, in a strong coupling analysis, rather than the mass, the  magnitude of the coupling constant~$\lambda / \alpha^4$ of the canonically normalized  field~$\varphi_{\rm c}:=\alpha \varphi$ is  the relevant quantity.
In particular, in an mGR context, we can  be confident of the characteristic analysis  as long as the energy scale is larger than the---\emph{observationally very small}---graviton mass. This energy scale can  therefore be \emph{much smaller} than the strong coupling scale~$\left(m^2M_{\rm P}\right)^{1/3}$\cite{deRham:2014zqa}. Of course, in models (such as mGR) with constraints, the 
mass terms can re-enter the characteristic matrix via the constraints themselves. Indeed it has been suggested in~\cite{Georgi} that characteristic surfaces indicate a breakdown of the theory's effective description of some putative ultraviolet completion. This viewpoint has been questioned recently in~\cite{Kaloper:2014vqa}.

\subsection{Characteristics and Branches}\label{trees}

The characteristic method neatly captures the difficulties faced by constrained systems whose solutions have multiple branches. As an example,
consider the following constraint equation,
\begin{eqnarray}
C[\phi^{(a)}\big] =C_1\big[\phi^{(a)}\big]\, C_2\big[\phi^{(a)}\big]=0,
\label{C1C2}
\end{eqnarray}
where~$C_1$ and~$C_2$ are independent functionals of dynamical fields~$\phi^{(a)}$.
Here we have two obvious choices for  solving the constraint~$C=0$, namely~$C_1=0$ or~$C_2=0$. Each choice yields a different branch of solutions. 
Suppose we choose the constraint~$C_1=0$. 
Although Eq.~(\ref{C1C2}) is not  quasi-linear, we can act with~$\xi^\mu\partial_\mu$ to yield a, first order,  quasi-linear equation:
\begin{eqnarray*}
C_2[\phi^{(a)}]\xi^\mu\partial_\mu C_1[\phi^{(a)}]+C_1[\phi^{(a)}]\xi^\mu\partial_\mu C_2[\phi^{(a)}]=0\, .
\label{pC1C2}
\end{eqnarray*}
 Given the  choice~$C_1=0$,
the second term vanishes, so 
the corresponding characteristic equation is~$C_2=0$. 
If we change branch from~$C_1=0$ to~$C_2=0$ along some hypersurface, then this surface must be characteristic.
Thus, upon branch change, evolution ceases to be unique. 
In a Hamiltonian analysis, this pathology  manifests itself  as a  singular structure of the Dirac bracket. 

\section{Fiducial massive gravity}\label{fmGR}


Fiducial mGR (fmGR) consists of a dynamical massive spin-2 field on a fixed, fiducial background~$\bar g_{\mu\nu}$, 
free of the bulk ghost excitations that  plague mGR theories~\cite{dRGT1}. Models of this type were first uncovered~\cite{Zumino}
in a study of Fierz--Pauli limits of one the metrics in the original ``$f$-$g$'' two metric model~\cite{Salam}, keeping the other as a fixed background. However the whole subject  lay  dormant  
after it was discovered~\cite{BD} that, at least for generic mass terms, 
a sixth, ghost, excitation appeared beyond linear order. 
It then took forty years until effective field theory decoupling limit technology was employed to show that a distinguished, three-parameter family of mass terms might be
ghost-free~\cite{dRGT1}. This was verified for generic regions of field space at full non-linear order in~\cite{HR}. 
Those studies were performed in a, rather intricate, metric formulation; later, it was realized that the original, highly efficient,  frame-like methods of~\cite{Salam}
could be employed~\cite{DeffayetBein,Chamseddine:2011mu, Chamseddine:2011bu, Hinterbichler:2012cn, Hassan:2012wt, Nomura:2012xr}. 
This allowed a covariant analysis of the models' degrees of freedom~\cite{DMZ,DW,DSW} and its characteristic matrix~\cite{DW,DSW,Izumi:2013poa,Deser:2013eua}.
Recently these analyses were extended to the models' full parameter range~\cite{DSWZ}; the key advantage of the frame-like approach  was that differential forms 
are exactly the correct objects for covariantly describing the hypersurfaces required for evolution and constraint analyses. In that language, suppressing wedge products whenever obvious, the action is simply
\begin{equation}
\begin{split}
S[e,\omega]= - \epsilon_{mnrs} \int \,  &\left(\frac14\, e^m{{}} e^n \, \left[d\omega^{rs} + \omega^r{}_t{{}}\omega^{ts}\right]\right.\\
&\left.\ \ -m^2 e{}^m{{}}\left[\frac{\beta_{0}}{4}e^n{{}} e^r{{}} e^s +\frac{\beta_{1}}{3}e^n{{}} e^r{{}} f^s+\frac{\beta_{2}}{2}e^n{{}} f^r{{}} f^s+\beta_{3}f^n{{}} f^r{{}} f^s\right]\right)\, .
\end{split}
\end{equation}
Here~$e^m:=e_\mu{}^m dx^\mu$,~$e_\mu{}^m \eta_{mn} e_\nu{}^n =g_{\mu\nu}$ and similarly for the fiducial variables.
For the model to have a linear Fierz--Pauli limit about its fiducial background, its parameters must obey
\begin{equation}\label{linearization}
m^2\left(\beta_0+\beta_1+\beta_2+\beta_3\right)=\frac{\bar\Lambda}6\, ,
\end{equation}
where~$\bar \Lambda$ is the fiducial cosmological constant. In that case, the Fierz--Pauli mass is
\begin{equation}\label{mFP}
m^2_{\rm FP}=m^2\left(\beta_1+2\beta_2+3\beta_3\right)\, .
\end{equation}

\subsection{Constraints}\label{Constraints}

The models' constraints are very easily understood; here is their counting, as given in~\cite{DSWZ}:
\begin{enumerate}
\item[0.] There are forty equations of motion. Varying the independent spin connection~$\omega^{mn}$ imposes 
vanishing torsion and varying the vierbein~$e^m$ sets the Einstein tensor equal to the stress tensor of the mass terms.
\item[1.] Since the equations of motion are forms, evaluating them along a spacelike hypersurface cannot give any time derivatives on 
dynamical fields; this gives sixteen primary constraints.
\item[2.]  The anti-symmetric part of the Einstein tensor, plus the covariant curl of the Einstein tensor equation of motion yield the so-called 
symmetry and vector constraints. These constitute ten secondary constraints.
\item[3.] The covariant curls of the symmetry and vector constraints yield the four remaining tertiary constraints. 
\end{enumerate}
The above constraint structure leaves ten independent variables subject to ten independent first order evolution equations, and thus five physical degrees of freedom.

\section{fmGR Characteristics}\label{char}

Since we are interested in vanishing of the characteristic matrix's determinant, our strategy is to use Gau\ss ian elimination to maximally simplify its associated homogeneous
linear equations, and then study the determinant of the remaining, {\it reduced characteristic matrix} determined this way.
We compute the shocks in the equations of motion and constraints to find  this matrix. For that, one begins by writing the equations of motion in first order form
\begin{eqnarray}
0&=&\nabla e^m\, ,\nonumber\\[1mm]
0&=& \frac12 \epsilon_{mnrs} e^nR^{rs} -m^2 t_m\, ,\nonumber\\[1mm]
0&=& d \omega^{mn} + \omega^m{}_t \omega^{tn}-R^{mn}\, . \label{PDEs}
\end{eqnarray}
Here we have treated the Riemann curvature two-form as an independent field and therefore added the third equation above to the system.
Also,~$\nabla$ is the covariant curl with respect to the spin connection~$\omega^{mn}$ and the mass stress tensor is encoded by the three-form
$$
t_m:=\epsilon_{mnrs}\big[\beta_0 e^n e^r e^s +
\beta_1 e^n e^r f^s +
\beta_2 e^n f^r f^s +
\beta_3 f^n f^r f^s   \big]\, .
$$
Then we denote the shock discontinuity across~$\Sigma$ (see~Eq.~(\ref{jump})) in~$(e^m, \omega^{mn},R^{mn})$ by
$$
[\partial_\mu e^m]=\xi_\mu \tilde e^m\, ,\quad
[\partial_\mu \omega^{mn}]=\xi_\mu \tilde \omega^{mn}\, ,\quad
[\partial_\mu e^m]=\xi_\mu \tilde R^{mn}\, . 
$$
To study characteristics that are spacelike with respect to the dynamical metric, we assume (without loss of generality), along the shock hypersurface~$\Sigma$, that
$$
\xi_\mu g^{\mu\nu}\xi_\nu= -1\, .
$$
Thus we can use the projector~$\Pi^\mu_\nu:=\delta^\mu_\nu+\xi^\mu\xi_\nu$ to decompose tensors, and hence any differential form~$\alpha$, as
$$
\alpha:=  \xi\,  \ring{\alpha} + \bm \alpha
$$
where~$\xi:=\xi_\mu dx^\mu$ and~$\bm \alpha$ is the purely spatial part of~$\alpha$.  In this notation, we can easily compute the shocks in the PDEs in Eq.~(\ref{PDEs}) as well as that of the covariant curl of the third of these (which says~$\nabla R^{mn}=0$). Altogether, this forces vanishing of the spatial parts of the shock profiles~$(\tilde e^m,\tilde \omega^{mn},\tilde R^{mn})$:
$$
\tilde {\bm e}^m=0=\tilde {\bm \omega}^{mn}=\tilde{\bm  R}{}^{mn}\, .
$$
If we restrict ourselves to the branch where the symmetry constraint (recall that symmetry of the Einstein tensor implies the same for the mass stress tensor) is solved via
$$
f^m e_m=0\, ,
$$
then the analysis is simplified by using~$f_{\mu\nu}=f_{\mu}{}^m e_{\nu m}$ as independent variable. The shock of the symmetry constraint implies that 
$$
\tilde e^m = \xi\,  \xi^\mu l_\mu{}^m \tilde f_{oo}
$$
where
$$
\tilde f_{oo}:=\xi^\mu \xi^\nu \tilde e_{\mu}{}^m f_{\nu m}\, ,
$$
and~$\l^\mu{}_m$ denotes the inverse fiducial vierbein. 

At this point, the  only independent shock profiles with timelike components, namely~$(\tilde f_{oo}, \ring{\tilde \omega}{}^{mn},\ring{\tilde R\, }{}^{mn})$, remain.
The shock in the curvature identity (which becomes an equation of motion now that the Riemann curvature is taken independent) 
$$
\nabla^\mu R_{\mu\nu\rho\sigma}=2\nabla_{[\rho} G_{\sigma]\nu} +g_{\nu[\rho}\nabla_{\sigma]} G_\mu{}^\mu\, ,
$$
can be used to determine the profiles~$\ring{\tilde R\, }{}^{mn}$ (see~\cite{DSWZ} for the detailed formula which will not be important to us here). This leaves seven
shock profiles~$(\tilde f_{oo}, \ring{\tilde \omega}{}^{mn})$ undetermined. They obey a set of homogeneous linear equations that determine the reduced characteristic
matrix. These are determined from the shocks of the vector, scalar and curled symmetry constraints. Schematically the result takes the form
$$
\begin{pmatrix}
M\: &N_{mn}\\[1mm]P_{\phantom{mn}}^{(6)}&Q^{(6)}_{mn}
\end{pmatrix}
\begin{pmatrix}
\tilde f_{oo}\\[1mm]\ring{\tilde \omega}{}^{mn}
\end{pmatrix}=0\, .
$$
The label ``$(6)$'' is used to indicate a column vector built from a pair of (spatial) three-vectors, so the reduced characteristic matrix is~$7\times 7$.
It depends on (i)  the dynamical and fiducial  vierbeine, (ii)  the {\it contorsion} tensor given by the difference between dynamical and fiducial connections, (iii) the  fiducial curvature,  and (iv) for models with~$\beta_3\neq 0$, the dynamical curvature.

To ensure the absence of superluminal shocks, one would need to demonstrate that the reduced characteristic matrix determinant cannot vanish anywhere.
Clearly, this is highly unlikely, given its dependence on both dynamical and fiducial fields. However, there does remain the hope that, at least for
some magical  combination of parameters  (and possibly a special choice of fiducial background), this miracle occurs. Not only is there no evidence
for this, but  counterexamples are rather easy to find, as we shall soon show. Superluminal shocks are therefore generic features of fmGR models.

The above argument can also be adapted to show that there are superluminal shocks with respect to the fiducial metric; this implies that the model, viewed as
a quantum theory of a massive spin two field propagating in a background, is pathological. However, there remains the possibility that it somehow still  manages to avoid closed timelike curves (CTCs)---the presence of superluminal propagation is, after all, just a warning that something nasty may plague the theory.  Indeed, as pointed out in~\cite{Geroch1}, superluminality may not always present a problem since it does not necessarily lead to acausality~\cite{Andreka:2014uya} (the existence of closed causal curves~\cite{CH}; see~\cite{Nester8,Ong:2013qja} for examples of acausality in modified theories of gravity without diffeomorphism invariance or local Lorentz invariance).  Here, on the contrary however, we will present an \emph{explicit} construction of closed causal curves\footnote{Here we differentiate (slightly pedantically) between CTCs---closed curves in a spacetime whose tangent vector is everywhere timelike---and CCCs---loops around which  initial data can be transmitted back to its starting point (typically by propagating shocks around loops along characteristic surfaces).} (CCCs).

\subsection{Acausality}

First note that
acausality here refers to local (infinitesimal) closed propagation curves. This ``micro-acausality''  differs from acausal~GR solutions: the (G\"odel) CTCs in~GR are \emph{global}, in the sense that locally in a neighborhood of a point on the curve, an observer can always define future and past. 
In contrast, the fmGR acausalities are \emph{local} so that  the causal structure is broken even in an infinitesimal region.
Indeed, CCCs in fmGR are \emph{generated dynamically}, whereas in~GR, starting from a spacetime without closed causal curves, it is at best extremely difficult, if not impossible to form  CCCs  without breaking energy conditions or causing  event horizons that  shroud the CCCs~\cite{DJtH}.

To construct an explicit example of acausality, we
consider the case where the  background and dynamical metrics (the latter viewed as a mean field along which  shocks propagate) are both flat
but not causally aligned:
\begin{equation}\label{flatonflat}
 d\bar s^2 = -dz^2 +dx^2+dy^2+dt^2\quad \mbox{and}\quad ds^2 = -dt^2+dx^2 +dy^2 +dz^2\, ,
\end{equation}
with
$f^3=e^0=dt$,~$f^1=e^1=dx$,~$f^2=e^2=dy$ and~$f^0=e^3=dz$.
This configuration is an fmGR solution so long as (in addition to the linearization condition Eq.~(\ref{linearization}) at~$\bar \Lambda =0$)  its parameters obey~\cite{DSWZ}
\begin{equation}\label{tune}
\beta_1+2\beta_2+3\beta_3=0\, .
\end{equation}
Notice that this requirement implies that the linearized FP mass~$m_{\rm FP}=0$ (see Eq.~(\ref{mFP})), so one might already rule this model out on the basis that its non-linear and linearized DoF counts do not agree. Nonetheless, it exemplifies the generic difficulties faced by models with a  field-dependent characteristic matrix. In the current ``flat on flat'', and hence contorsion-free, setting, the reduced characteristic matrix simplifies significantly to
\begin{eqnarray}
\begin{pmatrix}
\ M\qquad &0\\[2mm]
\ 0\qquad&\bm e^m \wedge \bm f^n\\[2mm]
\ 0\qquad&\epsilon^{mn}{}_{rs}\big[\beta_1 \bm e^r \wedge \bm e^s
+2\beta_2 \bm e^r \wedge \bm f^s
+3\beta_3 \bm f^r \wedge \bm f^s
\big]
\end{pmatrix}
\begin{pmatrix}
\tilde f_{oo}\\[1mm]\ring{\tilde \omega}{}_{mn}
\end{pmatrix}=0\, . \label{cha}
\end{eqnarray}
Here, because this matrix is block-diagonal, the exact form of the scalar~$M$ will not concern us; it is given in~\cite{DSWZ}. Note that because~$\bm e^m\wedge \bm f^n$ is a spatial two-form, contracted on the shock profile~$\ring{\tilde\omega}{}_{mn}$, this gives three conditions, and the same comment applies for the last line of the matrix above, so we have, as promised, a~$7\times 7$ homogeneous linear system of equations. Now consider the two, relatively tilted,  timelike vectors
$$
\xi_{A}=\frac\partial{\partial t} \quad \mbox{and}\quad \xi_B=\frac\partial{\partial t}+\alpha\frac\partial{\partial z}\, ,\quad (0<\alpha<1)\, .
$$
Vector A defines spacelike hypersurfaces~$t=$constant, which we call~$\Sigma_A$. These are characteristic\footnote{Here will use this property to send signals around a CTC using superluminal shocks. We also now know that  initial data along~$\Sigma_A$  (determined by restriction of the solution Eq.~(\ref{flatonflat}) to one of the characteristics surfaces~$\Sigma_A$) does not uniquely determine subsequent evolution. Note  that this does NOT mean that there are no solutions to the original PDEs with this initial data: rather, there are many of them, one of which is Eq.~(\ref{flatonflat}). In other words, the model has lost predictive power.} because the shock  profile condition imposed by the second line of the reduced characteristic matrix (calling $\xi^\mu \omega_{\mu mn}:=\omega_{omn}$)
$$\bm e^m \wedge \bm f^n\, \tilde\omega_{omn}=
2\tilde\omega_{o12}\, dx\wedge dy\,  
+
\big([\tilde\omega_{o10}+\tilde\omega_{o13}]dx
+[\tilde\omega_{o20}+\tilde\omega_{o23}]dy\big)\wedge dz\, ,
$$
has solutions for non-vanishing shock profiles~$\tilde\omega_{o01}=\tilde\omega_{o13}$,~$\tilde\omega_{o02}=\tilde\omega_{o23}$ and~$\tilde\omega_{o12}=0$,
along which the condition implied by the third line of the reduced characteristic matrix is trivially satisfied (so long as the parameters obey the on-shell condition Eq.~(\ref{tune})). 
We now need to find a second, independent, family of spacelike hypersurfaces,~$\Sigma_B$ (say), since we could then form a CTC by sequentially sending (i) a right-moving\footnote{This  slight abuse of language is designed to help the reader visualize the~$(t,z)$ projection of Minkowski spacetime.} signal along~$\Sigma_A$, (ii) a right-moving signal along~$\Sigma_B$, (iii) a left-moving signal along~$\Sigma_A$, (iv) a left-moving signal along~$\Sigma_B$ back to the starting spacetime point.

Thus it only remains to establish that surfaces of constant\footnote{Notice~$d\tau=(-1,0,0,\alpha)_\mu{} dx^\mu$ so~$\xi^\mu\partial_\mu= (-1,0,0,1)_\mu \eta^{\mu\nu}\partial_\nu=\frac\partial{\partial t}+\alpha\frac\partial{\partial z}=\xi_B$. We do not bother normalizing the vector~$\xi_B$ since this only contributes an overall irrelevant constant.}~$\tau=-t+\alpha z$ are characteristic.
For that we recompute the condition given by second line of the reduced characteristic matrix along~$\Sigma_A$ and now find
$$\bm e^m \wedge \bm f^n\, \tilde\omega_{omn}=
2\tilde\omega_{o12}\, dx\wedge dy\,  
+
(1+\alpha)\big([\tilde\omega_{o10}+\tilde\omega_{o13}]dx
+[\tilde\omega_{o20}+\tilde\omega_{o23}]dy\big)\wedge dz\, ,
$$
which is solved exactly as above. The third line of the characteristic equation again gives no new conditions, so~$\Sigma_B$ is characteristic (for any choice of tilting~$\alpha$).


It is not difficult to generate even  more general acausalities valid for any choice of the model's parameters $(\beta_0,\beta_1,\beta_2,\beta_3)$: 
 For simplicity, we again consider a flat  fiducial metric with tetrads 
$f^0=dt$, $f^1=dx$, $f^2=dy$, $f^3=dz$ for which  the fiducial connection  vanishes.
Now suppose that,  along  a constant-$t$ hypersurface~$\Sigma$, the fiducial and the dynamical tetrads are aligned but have different amplitudes:
\begin{eqnarray*}
e^0 = A\,dt ,\quad e^1 = B\,dx  \quad e^2 = C\,dy, \quad e^3 = -B\,dz \, ,
\end{eqnarray*}
where $A$, $B$ and  $C$ are constants. Observe, for later, that 
$e^1$ and $e^3$ have the same amplitudes but  different signs.
We further assume that  all connection components, as well as their spatial derivatives  vanish along this ``initial'' hypersurface~$\Sigma$.
Thus, along~$\Sigma$, the only non-vanishing derivatives of dynamical fields are with respect to the fiducial time coordinate~$t$.
The initial configuration for the Riemann tensor (when viewed as an independent field) is determined directly from the third, algebraic, equation in Display~(\ref{PDEs}).

We must first check whether this initial configuration is 
consistent with the constraints (these  were given explicitly for the first time in~\cite{DSWZ} and are summarized in Section~\ref{Constraints} above):
\begin{enumerate}
\item[(i)] The sixteen primary constraints follow from the spatial parts 
of the equations in Display~(\ref{PDEs}). Only the second of these gives a non-trivial condition:
$$
0=2m^2\left(-3\beta_0\, B^2C-\beta_1\, B^2+\beta_2 \, C+3\beta_3\, \right).
$$
This condition can be satisfied by 
tuning~$C$ to
\begin{eqnarray}
C=\frac{3\beta_3-\beta_1B^2}{3\beta_0B^2-\beta_2}\, .\nonumber
\end{eqnarray}
The singular cases $C=0,\infty$ are avoided by requiring that 
the parameter $B$ obeys neither  $3\beta_3-\beta_1B^2=0$ nor $3\beta_0B^2-\beta_2=0$.
\item[(ii)]
The ten secondary constraints are both trivially satisfied. The six symmetry constraints require that $e^m f_m=0$ along the initial surface which obviously holds, while the four vector constraints are proportional to the contorsion which vanishes along the~$\Sigma$ for this configuration (this need not be the case upon evolving the system).
\item[(iii)] Of the four remaining tertiary constraints, three are proportional to the contorsion (they are just the covariant curl of the secondary symmetry constraint). Thus only the scalar constraint must be verified. As shown in~\cite{DSWZ} (see their Equation~(15), Section~2.3.2), for vanishing contorsion and  flat fiducial backgrounds, this constraint is rather simple
$$
\epsilon_{mnrs}\big(\beta_1 e^m e^n+2\beta_2 e^m f^n +3\beta_3 f^m f^n \big)\wedge R^{rs}\approx 0\, .
$$
Along~$\Sigma$ the only non-vanishing part of the Riemann tensor is  $dt\wedge dx^j R_{tj}{}^{rs}$ so the sums over $m$ and $n$ in the above display can run only over values $1,2,3$ (because $dt\wedge dt =0$) so the scalar constraint takes the form $$M_{12} dx\wedge dy \wedge R^{03}+M_{23} dy\wedge dz \wedge R^{01}+
M_{31} dz\wedge dx \wedge R^{02}\approx 0\, ,$$
for constants $M_{ij}$.
 Thus only the part of $R^{0i}\propto dt\wedge dx^i$ can contribute to the scalar constraint. But these can be computed from the second equation in Display~(\ref{PDEs}) (note that this feature is guaranteed by the covariant proof given in~\cite{DSWZ} that the above is a constraint). This gives a (somewhat complicated) expression for the scalar constraint which is linear in the variable $A$. The key point is, that for {\it any} choice of parameters $(\beta_0,\beta_1,\beta_2,\beta_3)$, a value for $A$ can be found such that the scalar constraint holds. Note that the constant $B$ is still free, so we have really found a one parameter family of field data for which $\Sigma$ will be characteristic.
\end{enumerate}

 Next we demonstrate that the initial hypersurface~$\Sigma$ is 
 characteristic. Because the contorsion vanishes along~$\Sigma$, we can use the simplified version of the characteristic matrix given in Eq.~(\ref{cha}). The key feature of this initial configuration is that the coefficients of $e^1$ and $e^3$ have the same amplitude but opposite signs.
 Then, the $dx\wedge dz$ component in the middle line of the reduced characteristic matrix gives no condition. Thus, the hypersurface $\Sigma$ is characteristic.

We are now ready to generate  acausalities. 
The above analysis shows that, in a flat Minkowski background there are configurations along constant $t$ initial hypersurfaces whose evolution is not uniquely determined. This implies 
superluminal (with respect to the background) propagation along this characteristic hypersurface. To generate propagation around local loops, we
first  note that the above configuration  has a parity symmetry:  
the action is invariant under a simultaneous  flip of pairs of fiducial and dynamical tetrad components, for example $(e^1,f^1,e^2,f^2) \to (-e^1,-f^1,-e^2,-f^2)$.
(The mass term is obviously invariant under this transformation while invariance of the Einstein--Hilbert term requires also transforming  $\omega^{12}\to\omega^{12}$, $\omega^{xy}\to \omega^{xy}$, $\omega^{1x}\to -\omega^{1x}$ and $\omega^{2x}\to -\omega^{2x}$ where $x,y\neq 1,2$.) 
For our initial configuration, the map $(e^1,f^1,e^2,f^2) \to (-e^1,-f^1,-e^2,-f^2)$ produces a new solution to the constraints (because changing coordinates $(x,y)\to (-x,-y)$ yields the original configuration back again). In this way we use parity symmetries to generate new solutions. 
Now consider a small signal propagating in the direction $(0,a,b,c)$.
The parity symmetry can be used to generate signals propagating also in the directions $(0,a,-b,-c)$, $(0,-a,b,-c)$ and $(0,a,-b,-c)$. 
From these, a CCC can clearly be constructed.

The above examples are damning for fmGR, but yeasayers might complain that these are only very special initial conditions. Needless to say, however, one counterexample suffices to establish inconsistency. Actually the real lesson is that generating ill-posed configurations is easy, and likely a completely generic feature of the model.
In any case, what might seem to be a possible (albeit unlikely) escape route---careful tunings of the parameters
and fiducial background designed to avoid spacelike characteristics---is actually a major shortcoming;
 absent some guiding principle for choosing the fiducial space,  observational predictability seems to be completely lost in the swamp of what amounts to infinitely many 
tunable parameters.    Indeed,
 another essential ingredient in completing these models is to
specify, and see the consequences of, their coupling to matter~\cite{Dagain}. For fmGR,
a variety of difficulties were already exhibited in~\cite{Deser:2013rxa} (see also~\cite{Gullu:2013yha}); these were unavoidably and precisely  linked  to their constraint structure. In a theory with two
 metrics it is unclear how to couple matter's stress energy without destroying biGR's physical DoF content. [See, for example, the conflicting works~\cite{Yamashita:2014fga} and~\cite{deRham:2014naa} (as well as~\cite{Noller}) regarding whether any two-metric combinations can consistently couple to matter.]  
Thus, attempting to leave the borders
of~GR 
  forces one ineluctably back to it and the  gauge principle as the unique mechanisms for coupling geometry to matter.

\section{Outlook: Bimetric Gravity?}\label{conc}

The necessary incompatibility of the dynamical and fiducial causal structures and its attendant superluminality and acausalities, as well as the massive loss of observational predictability implied by choosing, by fiat, a fiducial metric, leaves biGR as the only conceivable  resolution:
perhaps a magical combination of  two dynamical metrics could determine a good causal structure.
While there is as yet no complete study of biGR's characteristics,
there is already strong evidence against its consistency. This was garnered by considering a putative partially massless (PM) limit. Recall that for the linear FP theory in  cosmological backgrounds, there is an intermediate spin~2 theory, whose excitations were lightlilke, but describe  only helicities~$(\pm2,\pm1)$~\cite{PM,PM1}. The PM model was discovered by tuning the FP mass to the cosmological constant in order to achieve gauge invariant,  lightcone propagation. Both fmGR and biGR possess PM free limits for special points in their parameter spaces~\cite{mGRpm,HassanPM}. However, both these models suffer from a new version of the original ``fatal flaw''~\cite{BD} of the generic mGR models, namely their interacting and linearized physical DoF contents do not match~\cite{DSW,deRhamPM,PMbi,PMbi1}. 
We hope to soon complete, and report on, the outcome of the explicit biGR consistency  calculations~\cite{DIOWZ}.

\section*{Acknowledgments} 

K.I. is
supported by Taiwan National Science Council under Project No. NSC101-2811-M-002-103.
A.W. thanks the Indian Institute of Science Bangalore, and the Harish-Chandra Institute, Allahabad for hospitality. S.D. was supported in part by NSF and DOE grants PHY-1266107 and  \#DE-SC0011632, respectively.

\appendix

\end{document}